\documentstyle[twoside,fleqn,espcrc2,psfig]{article}

\newcommand{\AmS}{{\protect\the\textfont2
  A\kern-.1667em\lower.5ex\hbox{M}\kern-.125emS}}

\title{Two-body spectra of pseudoscalar mesons 
       with an $O(a^2)$--improved lattice action 
       using Wilson fermions\thanks{Supported in part 
                                    by NSF PHY-9409195, 
                                    by OTKA T023844,
                                    by FWF P10468-PHY, 
                                    and by the Austrian 
                                    Ministry of Science (BMWFK)}}

\author{H. R. Fiebig\address{Physics Department, 
                            FIU-University Park, 
                            Miami, Florida 33199, USA},
        H. Markum\address{Institut f\"{u}r Kernphysik, 
                          Technische Universit\"{a}t Wien, 
                          A-1040 Vienna, Austria},
        A. Mih\'{a}ly\address{Department of Theoretical Physics, 
                              Lajos Kossuth University,
                              H-4010 Debrecen, Hungary},
        K. Rabitsch$^{\mbox{\scriptsize b}}$,
        and
        R. M. Woloshyn\address{TRIUMF, 
                              4004 Wesbrook Mall, 
                              Vancouver, BC, Canada, V6T 2A3}}

\begin{document}

\begin{abstract}
We extend our calculations with the second-order tree-level and
tadpole improved next-nearest-neighbor action to
meson-meson systems. Correlation matrices built from interpolating
fields representing two pseudoscalar mesons ($\pi$--$\pi$) with
relative momenta $\vec{p}$ are diagonalized, and the mass spectrum
is extracted. Link variable fuzzing and operator smearing 
at both sinks and sources is employed. Calculations are presented
for two values of the hopping parameter. 
The spectrum is used to discuss the
residual interaction in the meson-meson system.
\end{abstract}

\maketitle


The emergence of hadronic forces from first principles is a
fundamental question that poses itself to QCD. Remarkably,
only few attempts have been made to extract effective interactions,
or potentials, between two composite hadrons from the lattice 
\cite{firstattempts}.
This task is numerically very challenging since the residual
interaction between color singlet composites is about $10^{-2}$
to $10^{-3}$  smaller than a typical hadron mass.
In this paper we describe an attempt to study the residual
interaction in a simple meson-meson system from four-dimensional
lattice QCD.


The improved gluonic action used in our study 
includes planar six-link plaquettes
$U_{\mbox{\scriptsize rt}}$ in addition to the elementary
plaquettes $U_{\mbox{\scriptsize pl}}$
\begin{eqnarray}
S_{\mbox{\scriptsize G}}[U] &=& 
\beta\left[\sum_{\mbox{\scriptsize pl}}
(1-\frac13\mbox{Re}\mbox{Tr}U_{\mbox{\scriptsize pl}})\right .
\nonumber\\
&&
-\left .\frac{1}{20u_0^2} \sum_{\mbox{\scriptsize rt}}
(1-\frac13\mbox{Re}\mbox{Tr}U_{\mbox{\scriptsize rt}})\right] \ .
\end{eqnarray}
The improved Wilson fermionic action involves 
nearest-neighbor and next-nearest-neighbor couplings
\begin{eqnarray}
S_{\mbox{\scriptsize F}}[\psi,\bar{\psi};U] &=& \nonumber \\[1mm]
& & \hspace{-24mm} \sum_{x,\mu}\Big\{\frac{4}{3}\kappa\left[
\bar{\psi}(x)(1-\gamma_{\mu})U_{\mu}(x)\psi(x+\mu) \right.\nonumber \\
& & \hspace{-24mm} \left.+\bar{\psi}(x+\mu)
(1+\gamma_{\mu}) U_{\mu}^{\dagger}(x)\psi(x)\right] \nonumber \\[1mm]
& & \hspace{-24mm} -\frac{1}{6}\frac{\kappa}{u_0}
    \left[\bar{\psi}(x)(2-\gamma_{\mu})U_{\mu}(x)
    U_{\mu}(x+\mu)\psi(x+2\mu)
    \right. \nonumber \\[1mm]
& & \hspace{-24mm} \left.+\bar{\psi}(x+2\mu)(2+\gamma_{\mu})
U_{\mu}^{\dagger}(x+\mu)U_{\mu}^{\dagger}(x)\psi(x)\right]\Big\}
\nonumber \\[1mm]
& & \hspace{-24mm} -\sum_{x}\bar{\psi}(x)\psi(x)\ , 
\end{eqnarray}
with the hopping parameter $\kappa$.
Both actions are corrected for discretization errors to $O(a^{2})$ at the
classical level and contain the tadpole factor 
$u_{0}$ \cite{Egu84,Fie96d}.

Starting from a $\pi^{+}$-like field where the quarks carry the
flavors u and d
\begin{equation}
\phi_{\vec{p}}(t)=L^{-3}\sum_{\vec{x}}\,e^{i\vec{p}\cdot\vec{x}}
\bar{\psi}^{\mbox{\scriptsize d}}(\vec{x},t)\,\gamma_5
\psi^{\mbox{\scriptsize u}}(\vec{x},t)
\end{equation}
we construct a pion-pion operator \cite{Can97}
with relative lattice momentum $\vec{p}$
and total momentum $\vec{P}=\vec{0}$
\begin{equation}
\Phi_{\vec{p}}(t)=\phi_{-\vec{p}}(t)\,\phi_{+\vec{p}}(t)\ .
\end{equation}
To enhance the overlap of the interpolating field with the
ground state we employ link variable fuzzing 
and operator smearing 
\cite{Alb87}.
In particular, fuzzy links are used to
construct the smeared fermionic operators. Smearing is done
at both sinks and sources. Quark propagator matrix elements
are computed using a random-source technique \cite{Can97}.

The meson four-point time correlation matrix 
\begin{equation}
C_{\vec{p}\,\vec{q}}(t,t_0) =
\langle\Phi^{\dagger}_{\vec{p}}(t)\,\Phi_{\vec{q}}(t_0)\rangle-
\langle\Phi^{\dagger}_{\vec{p}}(t)\rangle
\langle\Phi^{\phantom{\dagger}}_{\vec{q}}(t_0)\rangle 
\end{equation}
consists of a free and an interaction contribution 
\begin{equation}
C=\overline{C}+C_{\mbox{\scriptsize I}} \ .
\end{equation}
The free two-pion correlator is diagonal
\begin{equation}
\overline{C}_{\vec{p}\,\vec{q}}(t,t_0) =
\delta_{\vec{p}\,\vec{q}}\,\,|c_{\vec{p}}(t,t_0)|^2 \ ,
\end{equation}
where $c_{\vec{p}}$ is the single-meson two-point function.
In the present study we are interested in the 
partial wave $\ell=0$ contained in the irreducible representation $A_1$
of the lattice symmetry group $O(3,{\cal Z})$.
We will identify the corresponding reduced matrix elements by
a superscript ${(A_1)}$.

The eigenvalues of the correlators decrease exponentially with
increasing time $t$
\begin{eqnarray}
\overline{C}^{(A_1)}(t,t_0) &\sim& e^{-\overline{E}_p(t-t_0)}
\quad\mbox{free} \nonumber \\
C^{(A_1)}(t,t_0) &\sim& e^{-E_n(t-t_0)} 
\quad\mbox{interacting} \ .
\label{Espec}
\end{eqnarray}
The mass spectra were obtained from linear fits to the logarithm of
the eigenvalues of the correlators.

We define an effective interaction through \cite{Can97}
\begin{equation}
H_{\mbox{\scriptsize I}} = -\frac{\partial}{\partial t} \ln 
(\overline{C}^{\,-1/2}\,C\phantom{x}\overline{C}^{\,-1/2}) \ .
\end{equation}
This yields the matrix elements 
$(\vec{p}\,|H_{\mbox{\scriptsize I}}|\vec{q}\,)$. 
The Fourier transform to coordinate space 
\begin{eqnarray}
(\vec{r}\,|H_{\mbox{\scriptsize I}}|\vec{s}\,)  &=&
L^{-3}\sum_{\vec{p}}\sum_{\vec{q}}
e^{ i\vec{p}\cdot(\vec{r}-\vec{s})}
e^{-i\vec{q}\cdot(\vec{r}+\vec{s})}\nonumber \\
& & 
(\vec{p}-\vec{q}\,|H_{\mbox{\scriptsize I}}|\vec{p}+\vec{q}\,) 
\nonumber \\[2mm]
 &=& \delta_{\vec{r},\vec{s}}\,V_{\mbox{\scriptsize loc}}(\vec{r}\,)+
\mbox{nonlocal part} 
\label{nonlocal}
\end{eqnarray}
contains a local potential
\begin{equation}
V_{\mbox{\scriptsize loc}}(\vec{r}\,) = \sum_{\vec{q}} e^{-2i\vec{q}\cdot\vec{r}}
(-\vec{q}\,|H_{\mbox{\scriptsize I}}|+\vec{q}\,) \ ,
\end{equation}
which stems from the $\vec{p}$ independent part of the $H_{\mbox{\scriptsize I}}$
matrix element in (\ref{nonlocal}).
The $s$-wave projection ($\ell = 0$) is
\begin{eqnarray}
V_{\mbox{\scriptsize loc}}(r) &=&
\frac{1}{4\pi}\int\,d\Omega_{\vec{r}}\,V_{\mbox{\scriptsize loc}}(\vec{r})\nonumber\\[1mm]
 &=& \sum_{q^2} j_0(2qr) (q^2|H_{\mbox{\scriptsize I}}^{(A_1)}|q^2) \ ,
\label{sVloc}
\end{eqnarray}
where $j_0$ is a spherical Bessel function.

\begin{figure*}[htb]
\centerline{\hbox{
            \hspace*{0.5mm}
            \psfig{figure=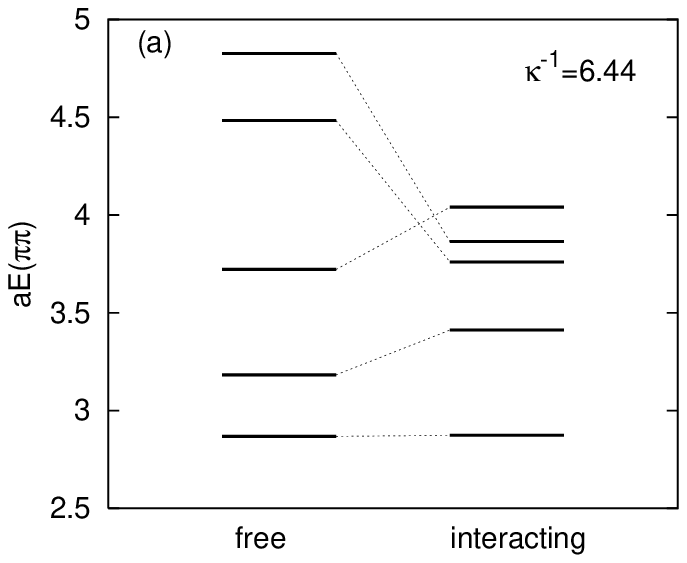}
            \hspace*{0.5mm}
            \psfig{figure=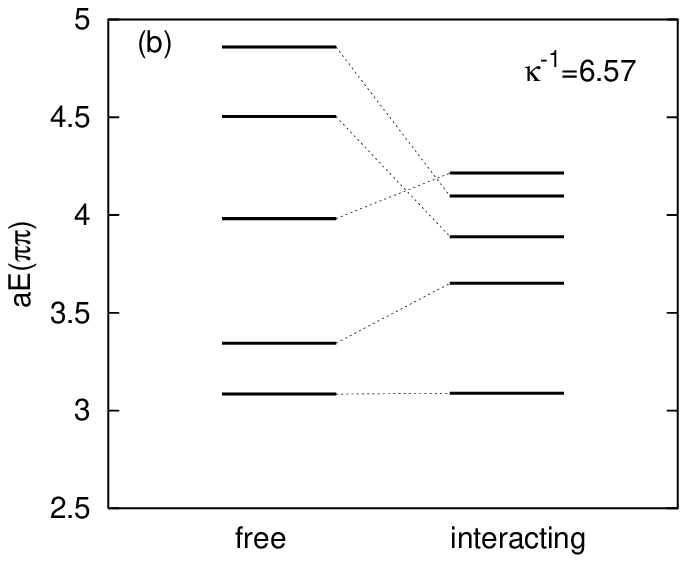}}}
\centerline{\hbox{
             \psfig{figure=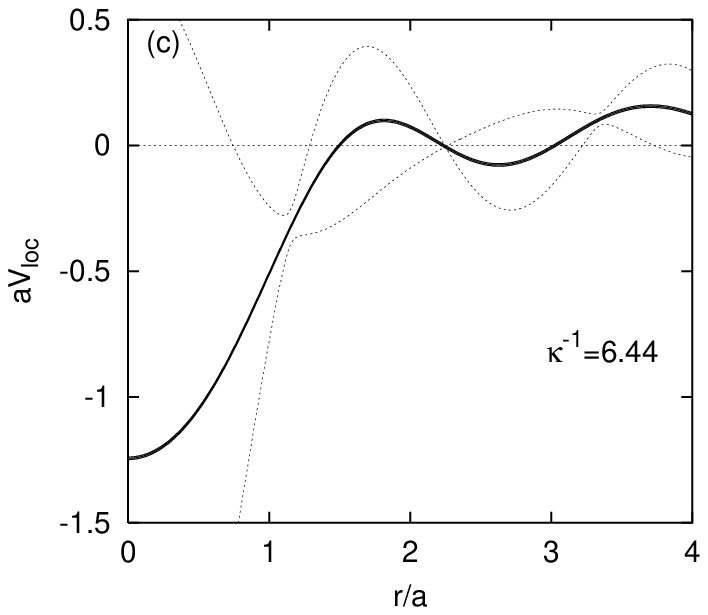}
             \psfig{figure=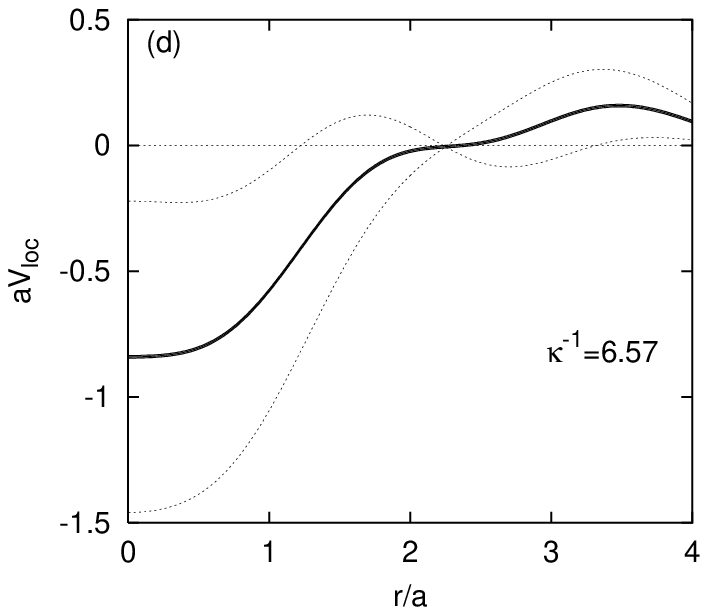}}}
\caption{Energy levels for the free and interacting pion-pion system for
$\kappa^{-1}=6.44$ (a) and $\kappa^{-1}=6.57$ (b) and the
corresponding $s$-wave projected effective local interaction potentials 
(c) and (d).}
\end{figure*}

This preliminary study of the two-pion system was performed on
an $L^3\times T = 9^3\times13$ lattice with
$\beta=6.2$ in the quenched approximation.
From the string tension the lattice constant is determined to be
$a\approx 0.4$fm corresponding to $a^{-1}\approx 500$MeV.
Measurements were taken on 48 independent gauge field
configurations separated by at least 1024 sweeps.
Figures 1 (a) and (b) show the energy spectra for the free and 
interacting two-pion system  according to (\ref{Espec})
for $\kappa^{-1}=6.44$ and $\kappa^{-1}=6.57$, respectively.
Since the critical value is $\kappa_C^{-1}\approx 5.5$ these
translate into rather large quark masses.
The spectra reveal that the residual interaction lowers the
large-momentum levels of the free system, thus indicating
attraction at short relative distances.

Figures 1 (c) and (d) display the $s$-wave projected local potential
resulting from (\ref{sVloc}). We obtain an attractive pion-pion
potential with a range of $<2a$ and a depth of about
$1a^{-1}$. 
A related study \cite{Mih97} with staggered fermions seems to give less
attraction.
The highest momentum $p=\frac{2\pi}{L}n$, $n=4$ caused some
numerical instability (though no qualitative change)
to $V_{\mbox{\scriptsize loc}}(r)$. We therefore only used $n=0\ldots 3$ in our
calculation of the effective local potential.
The solid lines in Figs.~1 (c) and (d) are the results using 48
gauge field configurations. The dotted lines represent the errors
from a jackknife analysis with 8 samples of 42 configurations each.
The large errors in (c) indicate 
that the results for $\kappa^{-1}=6.44$ are still numerically
unstable. 
For the current exploratory analysis no attempts were made to
estimate systematic errors.

We expect that a refined analysis, which would include extrapolation to the
chiral limit and proper isospin of the $\pi$--$\pi$ interpolating
fields, will make comparison to experimental findings possible.
Studies similar in spirit to the one initiated here may be extended to
various other hadron-hadron systems, like $\pi$--N for example.


\begin{thebibliography}{9}

\bibitem{firstattempts}
D.~G.~Richards, D.~K.~Sinclair, and D.~Sivers, Phys.~Rev. D 42 (1990) 3191;
K.~Rabitsch, H.~Markum, and W.~Sakuler, Phys.~Lett.  B 318 (1993) 507;
M. Fukugita, Y. Kuramashi, M. Okawa, H. Mino, and A. Ukawa,
Phys. Rev. D 52 (1995) 3003.

\bibitem{Egu84}
H. Hamber and C.M. Wu, Phys. Lett. B 133 (1983) 351;
T. Eguchi and N. Kawamoto, Nucl. Phys. B 237 (1984) 609.

\bibitem{Fie96d}
H.R. Fiebig and R.M. Woloshyn, Phys. Lett. B 385 (1996) 273.

\bibitem{Can97}
J.D. Canosa and H.R. Fiebig, Phys. Rev. D 55 (1997) 1487.

\bibitem{Alb87}
M. Albanese et al., Phys. Lett. B 192 (1987) 163;
%
C. Alexandrou,
S. G\"{u}sken, F. Jegerlehner, K. Schilling, and R. Sommer, 
Nucl Phys. B 414 (1994) 815.

\bibitem{Mih97}
A. Mih\'{a}ly, H.R. Fiebig, H. Markum, and K. Rabitsch, 
Phys. Rev. D 55 (1997) 3077.

\end{thebibliography}
\end{document}